\documentclass[prd,onecolumn,nofootinbib,showkeys]{revtex4}
\newcommand\gothfamily{\usefont{U}{ygoth}{m}{n}}
\DeclareTextFontCommand{\textgoth}{\gothfamily}

\begin{document}

\title{NONSINGULAR DIRAC PARTICLES IN SPACETIME WITH TORSION}

\author{{\bf Nikodem J. Pop{\l}awski}}

\affiliation{Department of Physics, Indiana University, Swain Hall West, 727 East Third Street, Bloomington, IN 47405, USA}
\email{nipoplaw@indiana.edu}

\noindent
{\em Physics Letters B}\\
Vol. {\bf 690}, No. 1 (2010) pp. 73--77\\
\copyright\,Elsevier B. V.
\vspace{0.4in}

\begin{abstract}
We use the Papapetrou method of multipole expansion to show that a Dirac field in the Einstein-Cartan-Kibble-Sciama (ECKS) theory of gravity cannot form singular configurations concentrated on one- or two-dimensional surfaces in spacetime.
Instead, such a field describes a nonsingular particle whose spatial dimension is at least on the order of its Cartan radius.
In particular, torsion modifies Burinskii's model of the Dirac electron as a Kerr-Newman singular ring of the Compton size, by replacing the ring with a toroidal structure with the outer radius of the Compton size and the inner radius of the Cartan size.
We conjecture that torsion produced by spin prevents the formation of singularities from matter composed of quarks and leptons.
We expect that the Cartan radius of an electron, $\sim10^{-27}\,$m, introduces an effective ultraviolet cutoff in quantum field theory for fermions in the ECKS spacetime.
We also estimate a maximum density of matter to be on the order of the corresponding Cartan density, $\sim10^{51}\,\mbox{kg}\,\mbox{m}^{-3}$, which gives a lower limit for black-hole masses $\sim10^{16}\,\mbox{kg}$.
This limit corresponds to energy $\sim10^{43}\,\mbox{GeV}$ which is 39 orders of magnitude larger than the maximum beam energy currently available at the LHC.
Thus, if torsion exists and the ECKS theory of gravity is correct, the LHC cannot produce micro black holes.
\end{abstract}


\maketitle

\section{Introduction}

Einstein's general theory of relativity (GR) is the geometric theory of gravitation which has been confirmed by many experimental and observational tests \cite{LL,MTW}.
However, this theory has a problematic feature: the appearance of curvature singularities, which are points in spacetime where the density of matter and curvature are infinite, and thus physics laws break down.
Such a singularity appears in the Kerr-Newman metric which is a solution of the Einstein-Maxwell equations describing a rotating, electrically charged mass \cite{KN}.
Remarkably, the Kerr-Newman solution has the same gyromagnetic ratio as that of the Dirac electron \cite{Car,KS}, which suggested treating this solution as a classical model of an extended electron in GR \cite{Isr}.
Recently, Burinskii has shown that the Dirac equation may be incorporated into the Kerr-Newman geometry in which the wave function of the Dirac electron acquires an extended spacetime structure: a singular ring of the Compton size \cite{Bur}.

The Einstein-Cartan or Einstein-Cartan-Kibble-Sciama (ECKS) theory of gravity naturally extends GR to include matter with intrinsic spin, which produces torsion, providing a more complete account of local gauge invariance with respect to the Poincar\'{e} group \cite{Lord,Hehl1,SH,Tr}.
It is a viable theory of gravity, which differs significantly from GR only at densities of matter much larger than the density of nuclear matter.
In this Letter we apply the Papapetrou method of multipole expansion to the conservation law for the spin density.
We show that this law, within the ECKS theory, prevents a Dirac field from forming singular configurations concentrated on one- (points or system of points) or two-dimensional (strings) surfaces in spacetime.
Instead, such a field forms nonsingular configurations whose spatial dimensions are at least on the order of its Cartan radius.
In particular, torsion modifies Burinskii's model by replacing a Dirac-Kerr-Newman ring singularity with a nonsingular toroidal structure with the outer radius of the Compton size and the inner radius of the Cartan size.
Consequently, we conjecture that torsion produced by spin eliminates the appearance of singularities from fermionic matter (quarks and leptons) which builds all stars.
We expect that the Cartan size of an electron introduces an effective ultraviolet cutoff in quantum field theory for fermions if the ECKS theory of gravity is correct.
We also estimate a maximum density of fermionic matter and a lower limit for black-hole masses in the ECKS theory.

\section{Einstein-Cartan theory}

In the ECKS theory of gravity, the tetrad $e^i_a$ and the spin connection $\omega^a_{\phantom{a}bk}=e^a_j(e^j_{\phantom{j}b,k}+\Gamma^{\,\,j}_{i\,k}e^i_b)$ are dynamical variables describing the geometry of spacetime \cite{Lord,Hehl1,SH,Tr}.
The symbol $_{,i}$ denotes differentiation with respect to $x^i$ and $\Gamma^{\,\,i}_{j\,k}$ is the affine connection.
The affine connection is asymmetric in the lower indices and its antisymmetric part is the torsion tensor, $S^i_{\phantom{i}jk}=\Gamma^{\,\,\,\,i}_{[j\,k]}$, where $[\,]$ denotes antisymmetrization.
The spin connection plays a role similar to $\Gamma^{\,\,i}_{j\,k}$; it appears in the covariant derivative of a Lorentz vector: $V^a_{\phantom{a};i}=V^a_{\phantom{a},i}+\omega^a_{\phantom{a}bi}V^b$.
The tetrad relates the spacetime coordinates $i,j,...$ to the local Lorentz coordinates $a,b,...$: $V^a=V^i e^a_i$.
The spacetime coordinates are lowered or raised by the metric tensor $g_{ik}$, as in GR, and the Lorentz coordinates by the Minkowski tensor of the special theory of relativity.
The variation of the Lagrangian density of matter $\mathcal{L}_m$ with respect to the tetrad defines the dynamical energy-momentum density: $\Theta^{\phantom{i}a}_i=\frac{\delta \mathcal{L}_m}{\delta e^i_a}$.
Its variation with respect to the spin connection defines the dynamical spin density, antisymmetric (like $\omega^{ab}_{\phantom{ab}i}$) in the indices $a,b$ \cite{Lord,Hehl1,SH}:
\begin{equation}
\Sigma_{ab}^{\phantom{ab}i}=2\frac{\delta \mathcal{L}_m}{\delta\omega^{ab}_{\phantom{ab}i}}.
\label{spden2}
\end{equation}
The invariance of $\mathcal{L}_m$ under Lorentz transformations (tetrad rotations) gives the conservation law for the spin density:
\begin{equation}
\Sigma^{ijk}_{\phantom{ijk},k}-\Gamma^{\,\,i}_{l\,k}\Sigma^{jlk}+\Gamma^{\,\,j}_{l\,k}\Sigma^{ilk}-2\Theta^{[ij]}=0.
\label{Pap22}
\end{equation}

The ECKS Lagrangian density \cite{Lord,Hehl1,SH} is given by
\begin{equation}
\mathcal{L}=\mathcal{L}_m-\frac{c^4}{16\pi G}{\sf e}R,
\label{ecks}
\end{equation}
where ${\sf e}=\mbox{det}\,e^a_i$, $R=R^b_{\phantom{b}j}e^j_b$ is the Ricci scalar (which is tetrad-rotation-invariant), $G$ is the gravitational constant, and $c$ is the speed of light.
The Ricci tensor $R^a_{\phantom{a}i}=R^{ab}_{\phantom{ab}ij}e^j_b$, where the curvature tensor $R^a_{\phantom{a}bij}=\omega^a_{\phantom{a}bj,i}-\omega^a_{\phantom{a}bi,j}+\omega^a_{\phantom{a}ci}\omega^c_{\phantom{c}bj}-\omega^a_{\phantom{a}cj}\omega^c_{\phantom{c}bi}$.
The curvature of spacetime is locally related to the energy-momentum density through the Einstein equations: ${\sf e}\biggl(R^a_{\phantom{a}i}-\frac{1}{2}Re^a_i\biggr)=\frac{8\pi G}{c^4} \Theta^{\phantom{i}a}_i$, which follow from the stationarity of the action corresponding to the ECKS Lagrangian density (\ref{ecks}) under variations of the tetrad.
The torsion of spacetime is locally related to the spin density (\ref{spden2}) through the Cartan equation:
\begin{equation}
{\sf e}(S^i_{\phantom{i}ab}-S_a e^i_b+S_b e^i_a)=-\frac{4\pi G}{c^4}\Sigma^{\phantom{ab}i}_{ab},
\label{Cartan}
\end{equation}
where $S_i=S^k_{\phantom{k}ik}$ is the torsion vector, which follows from the stationarity of the ECKS action under variations of the spin connection.
Combining the Einstein equations and (\ref{Cartan}) gives
\begin{equation}
G_{ik}=\frac{8\pi G}{c^4}T_{ik}+U_{ik},
\label{EC}
\end{equation}
where $G_{ik}=R_{ik}-\frac{1}{2}Rg_{ik}$ is the Einstein tensor of GR and $T_{ik}=\frac{2}{{\sf e}}\frac{\delta \mathcal{L}_m}{\delta g^{ik}}$ is the metric energy-momentum tensor \cite{LL}.
The tensor
\begin{equation}
U_{ik}=-(S^l_{\phantom{l}ij}+2S_{(ij)}^{\phantom{(ij)}l})(S^j_{\phantom{j}kl}+2S_{(kl)}^{\phantom{(kl)}j})+4S_i S_k+\frac{1}{2}g_{ik}(S^{mjl}+2S^{(jl)m})(S_{ljm}+2S_{(jm)l})-2g_{ik}S^j S_j,
\label{corr}
\end{equation}
where $(\,)$ denotes symmetrization, is quadratic in $\Sigma_{ij}^{\phantom{ij}k}$ \cite{Hehl1,dS,Hehl2}.
The ECKS Lagrangian density (\ref{ecks}) is the simplest one among various theories of gravity with torsion \cite{SH}.
However, the relation (\ref{Pap22}) is valid not only for the ECKS theory, but also for other theories of gravity with torsion produced by spin.
In GR, the torsion tensor vanishes, reducing (\ref{EC}) to the usual Einstein equations.

The Cartan equation (\ref{Cartan}) is a linear, algebraical relation; torsion is proportional to spin density.
Thus the torsion tensor vanishes outside material bodies, where the spin density is zero.
Unlike the curvature of spacetime, the torsion field in the ECKS theory does not propagate.
The appearance of torsion only inside material bodies introduces limitations on its detection and thus on experimental verification of the ECKS theory.
Typical experimental limits on torsion come from searches for dynamical properties such as: quantum effects from the coupling of torsion to the Dirac spinor, nongeodesic motion of Dirac particles in spacetime with torsion, neutron phase shifts in the presence of a coupling between the orbital angular momentum and torsion, forces from spin-spin interaction or Yukawa potential due to torsion, and corrections to the mass of a neutron star due to torsion \cite{SH,CF,test}.
The effects of torsion can also be indirect, such as a nonlinear character of the Dirac equation in the presence of torsion (due to the Heisenberg-Ivanenko term which is cubic in spinor fields \cite{dS}) and anomalies in the Standard Model in curved spacetime with torsion \cite{SH}.

\section{Papapetrou method for spin density}

Consider matter which is distributed over a small region in space and consists of points with the coordinates $x^i$, forming an extended body whose motion is represented by a world tube in spacetime.
The motion of the body as a whole is represented by an arbitrary timelike world line $\gamma$ inside the world tube, which consists of points with the coordinates $X^i(s)$, where $s$ is the affine parameter on $\gamma$.
Define
\begin{equation}
\delta x^\alpha=x^\alpha-X^\alpha,\,\,\,\delta x^0=0,\,\,\,u^i=\frac{dX^i}{ds},
\label{Pap1}
\end{equation}
where $\alpha$ denotes spatial coordinates.
The equations of motion for the body in general relativity, $u^i=u^i(s)$, result from the multipole expansion of the covariant conservation of the energy-momentum tensor, as shown by Mathisson and Papapetrou \cite{Pap}.
These equations were generalized by Nomura, Shirafuji and Hayashi to the Riemann-Cartan spacetime with torsion, and result from the covariant conservation laws for the spin density (\ref{Pap22}) and for the energy-momentum density \cite{NSH}.
They were also extended to a more general metric-affine gravity \cite{also}.

Define the following integrals \cite{Pap,NSH}:
\begin{eqnarray}
& & M^{ik}=u^0\int\Theta^{ik}dV, \label{Pap2} \\
& & M^{ijk}=-u^0\int\delta x^i\Theta^{jk}dV, \label{Pap3} \\
& & N^{ijk}=u^0\int\Sigma^{ijk}dV. \label{Pap4}
\end{eqnarray}
The quantity $N^{ijk}$ is a tensor.
The relation $\delta x^0=0$ in (\ref{Pap1}) gives
\begin{equation}
M^{0jk}=0.
\label{Pap6}
\end{equation}
Assume that the dimensions of the body are small, so integrals with two or more factors $\delta x^i$ multiplying $\Theta^{jk}$ and integrals with one or more factors $\delta x^i$ multiplying $\Sigma^{jkl}$ can be neglected.
Integrating (\ref{Pap22}) over the volume hypersurface and using Gau\ss-Stokes theorem to eliminate surface integrals gives
\begin{equation}
\int\Sigma^{ij0}_{\phantom{ij0},0}dV-\int\Gamma^{\,\,i}_{l\,k}\Sigma^{jlk}dV+\int\Gamma^{\,\,j}_{l\,k}\Sigma^{ilk}dV-2\int\Theta^{[ij]}dV=0.
\label{Pap23}
\end{equation}
The conservation law (\ref{Pap22}) also gives
\begin{equation}
(x^l\Sigma^{ijk})_{,k}=\Sigma^{ijl}+x^l\Gamma^{\,\,i}_{l\,k}\Sigma^{jlk}-x^l\Gamma^{\,\,j}_{l\,k}\Sigma^{ilk}+2x^l\Theta^{[ij]}.
\label{Pap25}
\end{equation}
Integrating (\ref{Pap25}) over the volume hypersurface and using Gau\ss-Stokes theorem to eliminate surface integrals gives
\begin{equation}
\int(x^l\Sigma^{ij0})_{,0}dV=\int\Sigma^{ijl}dV+\int x^l\Gamma^{\,\,i}_{m\,k}\Sigma^{jmk}dV-\int x^l\Gamma^{\,\,j}_{m\,k}\Sigma^{imk}dV+2\int x^l\Theta^{[ij]}dV.
\label{Pap26}
\end{equation}
Substituting (\ref{Pap1}) into (\ref{Pap26}) gives
\begin{eqnarray}
& & \frac{u^l}{u^0}\int\Sigma^{ij0}dV+X^l\int\Sigma^{ij0}_{\phantom{ij0},0}dV=\int\Sigma^{ijl}dV+2\int\delta x^l\Theta^{[ij]}dV \nonumber \\
& & +X^l(\int\Gamma^{\,\,i}_{m\,k}\Sigma^{jmk}dV-\int\Gamma^{\,\,j}_{m\,k}\Sigma^{imk}dV+2\int\Theta^{[ij]}dV\Bigr),
\end{eqnarray}
which reduces, due to (\ref{Pap23}), to
\begin{equation}
\frac{u^l}{u^0}\int\Sigma^{ij0}dV=\int\Sigma^{ijl}dV+2\int\delta x^l\Theta^{[ij]}dV
\end{equation}
or \cite{NSH}
\begin{equation}
M^{l[ij]}=-\frac{1}{2}\biggl(\frac{u^l}{u^0}N^{ij0}-N^{ijl}\biggr).
\label{Pap29}
\end{equation}
Putting $l=0$ in (\ref{Pap29}) gives the identity because of (\ref{Pap6}).

\section{Dirac field with torsion}

In relativistic quantum mechanics, an electron (or any other fermion) with mass $m$ and electric charge $q$, is described by a spinor field (wave function) \cite{IZ} which satisfies the Dirac equation: $i\gamma^i(\hbar\psi_{,i}+i\frac{q}{c}A_i\psi)-mc\psi=0$, where $A_i$ is the electromagnetic potential.
The $4\times4$ matrices $\gamma^i$ are the Dirac matrices which satisfy $\gamma^i\gamma^k+\gamma^k\gamma^i=2g^{ik}I$, where $I$ is the $4\times4$ unit matrix.
The Dirac equation results from the stationarity under variations $\delta\bar{\psi}$ of the action corresponding to the Dirac Lagrangian density: $\mathcal{L}_m=\mathcal{L}_\psi=\frac{i\hbar c}{2}(\bar{\psi}\gamma^i\psi_{,i}-\bar{\psi}_{,i}\gamma^i\psi)-q\bar{\psi}\gamma^i\psi A_i-mc^2\bar{\psi}\psi$, where $\bar{\psi}=\psi^\dag\gamma^0$ is the adjoint spinor and $\dag$ denotes the Hermitian conjugate.
In curved spacetime, the Dirac Lagrangian density becomes \cite{Lord,Hehl1,SH,dS}
\begin{equation}
\mathcal{L}_\psi=\frac{i\hbar c{\sf e}}{2}(\bar{\psi}\gamma^i\psi_{,i}-\bar{\psi}_{,i}\gamma^i\psi)-\frac{i\hbar c{\sf e}}{2}\bar{\psi}(\gamma^i\Gamma_i+\Gamma_i\gamma^i)\psi-q{\sf e}\bar{\psi}\gamma^i\psi A_i-mc^2{\sf e}\bar{\psi}\psi,
\label{sECg2}
\end{equation}
where the spinor connection $\Gamma_i$ is given by the Fock-Ivanenko coefficients: $\Gamma_i=-\frac{1}{4}\omega_{abi}\gamma^a \gamma^b$.
The spin density (\ref{spden2}) corresponding to the Lagrangian density (\ref{sECg2}) is totally antisymmetric \cite{Hehl1,dS,antis}:
\begin{equation}
\Sigma^{ijk}=\frac{i\hbar c{\sf e}}{2}\bar{\psi}\gamma^{[i}\gamma^j\gamma^{k]}\psi=\Sigma^{[ijk]}.
\label{sECg4}
\end{equation}
The definition (\ref{spden2}) implies that only the totally antisymmetric component of the torsion tensor couples to Dirac fields.
Moreover, the Cartan equation (\ref{Cartan}) yields the total antisymmetry of the torsion tensor, $S_{ijk}=S_{[ijk]}$.
The spin density (\ref{sECg4}) does not depend on $m$ and $q$, and it remains the same if we include the weak and strong interactions of fermions.
Substituting the spin density (\ref{sECg4}) into (\ref{sECg2}) introduces the Heisenberg-Ivanenko four-fermion self-interaction term in the Lagrangian density \cite{Hehl1,dS,Hehl2}:
\begin{equation}
\mathcal{L}_S=\frac{3\pi G{\sf e}}{2}(\hbar c)^2(\bar{\psi}\gamma^i\gamma^5\psi)(\bar{\psi}\gamma_i\gamma^5\psi).
\label{HI}
\end{equation}
The total antisymmetry of the spin density gives
\begin{equation}
N^{ijk}=N^{[ijk]}.
\label{antisym}
\end{equation}

Assume that a fermionic field forms the simplest configuration, i.e. a point (single-pole approximation).
For this configuration located at ${\bf r}$, $\Sigma^{ik}$ is proportional to ${\bm\delta}({\bf r})$ and does not contain derivatives of ${\bm\delta}({\bf r})$ (which do not represent any physical situation), so $M^{ik}$ is finite \cite{LL}.
We also have
\begin{equation}
M^{\alpha ij}\propto\int\delta x^\alpha u^{ij}\delta({\bf r})dV,
\end{equation}
where $u^{ij}$ is some finite tensor, which gives
\begin{equation}
M^{\alpha ij}=0.
\end{equation}
Thus (\ref{Pap29}) reduces to
\begin{equation}
N^{ijl}=\frac{u^l}{u^0}N^{ij0},
\label{eq1}
\end{equation}
which for a spinor field represented by the Dirac Lagrangian density (\ref{sECg2}) gives \cite{NSH}
\begin{equation}
N^{il0}=-\frac{u^l}{u^0}N^{i00}=0,
\label{eq2}
\end{equation}
due to (\ref{antisym}).
Reference \cite{NSH} concludes from (\ref{eq2}) that a Dirac particle in this approximation moves in the gravitational field like a spinless point particle, i.e. there is no corrections from spin to the (geodesic) equation of motion of such a particle.
However, substituting (\ref{eq2}) into (\ref{eq1}) gives
\begin{equation}
N^{ijk}=0.
\label{eq3}
\end{equation}
For a point particle located at ${\bf r}$, $\Sigma^{ijk}$ is proportional to ${\bm\delta}({\bf r})$, so (\ref{Pap4}) and (\ref{eq3}) imply
\begin{equation}
\Sigma^{ijk}=0,
\label{none}
\end{equation}
from which it follows (for the matter Lagrangian density given by (\ref{sECg2})) that $\psi=0$, i.e. there is no spinor field in the first place.
Equivalently, if $\psi\neq0$ then $M^{ijk}\neq0$, otherwise the conservation law for the spin density (\ref{Pap22}) and thus the invariance of the matter Lagrangian density under Lorentz transformations (tetrad rotations) would be violated.

Thus our conclusion is more fundamental: the single-pole approximation of a Dirac field is not a solution of the gravitational field equations in spacetime with torsion, not just a solution with a wrong equation of motion.
A Dirac field in the Riemann-Cartan spacetime of the ECKS theory must have $M^{ijk}\neq0$, so it cannot be a (singular) point distribution and thus it cannot represent a point particle.
Although this conclusion seems expected because higher moments should be included to encode the classical angular momentum, we confirm that these moments are also necessary to encode the intrinsic spin in order to obey the gravitational field equations.
Moreover, a Dirac field cannot be a system of points, like the pole-dipole approximation in \cite{HP,Cor}, because each such a point has the symmetric energy-momentum tensor for which (\ref{Pap3}) would give $M^{i[jk]}=0$.
In this case (\ref{Pap29}) and (\ref{antisym}) would still yield (\ref{none}) which contradicts (\ref{sECg2}) for $\psi\neq0$.

Now assume that a fermionic field forms a string.
From symmetry considerations we expect this string to be a ring, as a Kerr-Newman singularity \cite{Bur}.
The Kerr-Newman metric is a solution of the Einstein-Maxwell equations describing the field of an electrically charged mass $m$ (with charge $q$) rotating with angular momentum $J$ \cite{KN}.
This metric is given in the Boyer-Lindquist coordinates $(r,\theta,\phi)$ \cite{BL} by
\begin{eqnarray}
& & ds^2=\biggl(1-\frac{r_g r-r_q^2}{\rho^2}\biggr)c^2 dt^2-\biggl(r^2+a^2+a^2\mbox{sin}^2\theta\frac{r_g r-r_q^2}{\rho^2}\biggr)\mbox{sin}^2\theta\,d\phi^2 \nonumber \\
& & +2a\mbox{sin}^2\theta\frac{r_g r-r_q^2}{\rho^2}\,cdt\,d\phi-\frac{\rho^2}{\Delta}dr^2-\rho^2 d\theta^2,
\label{KeNe}
\end{eqnarray}
where $r_g=\frac{2Gm}{c^2}$, $a=\frac{J}{mc}$, $\rho^2=r^2+a^2\mbox{cos}^2\theta$, $r_q^2=\frac{Gq^2}{c^4}$ and $\Delta=r^2-r_g r+a^2+r_q^2$.
The Kretschmann scalar $R_{ijkl}R^{ijkl}$ for a Kerr-Newman field is singular for $\rho^2=0$ \cite{Hen}.
A coordinate transformation $x+iy=(r+ia)\mbox{sin}\theta e^{i\phi},\,z=r\mbox{cos}\theta$ to the Kerr-Schild coordinates $(x,y,z)$ \cite{KS} brings the Kerr-Newman metric to a form in which it tends at large $r$ to the Minkowski metric.
This form shows that a Kerr-Newman singularity is a ring of radius $a$ \cite{KN,Car}, which is consistent with M{\o}ller's theorem in the special theory of relativity stating that a system with a positive energy density, angular momentum $J$ and rest mass $m$ must have a finite extension $r>\frac{J}{mc}$ \cite{Mol}.

If a fermionic field, for which the angular momentum is equal to its spin, $J=\frac{\hbar}{2}$, forms a Dirac-Kerr-Newman ring then its radius $a=\frac{\hbar}{2mc}$ is on the order of the corresponding Compton wavelength \cite{Bur}.
This ring is a naked (without an event horizon) singularity because for all fermions $a^2+r^2_q>(\frac{r_g}{2})^2$ \cite{Car}.
From the symmetry considerations it follows that in the cylindrical coordinates $(x^1=r,x^2=z,x^3=\phi)$:
\begin{equation}
M^{\alpha ij}\propto\int\delta x^\alpha v^{ij}\delta(r-a)\delta(z)dr\,dz\,d\phi,
\end{equation}
where $\delta x^1=r-a$, $\delta x^3=z$ and $v^{ij}$ is some finite tensor.
Thus we have
\begin{equation}
M^{1ij}=M^{3ij}=0,
\end{equation}
for which (\ref{Pap29}) gives
\begin{equation}
N^{ij1}=\frac{u^1}{u^0}N^{ij0},\,\,\,N^{ij3}=\frac{u^3}{u^0}N^{ij0}.
\end{equation}
Substituting $j=0$ and using (\ref{antisym}) leads to
\begin{equation}
N^{i01}=N^{i03}=0,
\label{der}
\end{equation}
from which it follows that $N^{123}$ is the only nonzero component of $N^{ijk}$.
Thus the $z$-component of spin, $N_3$, which is dual to $N^{012}$, vanishes.
Consequently, a Dirac field in the ECKS gravity cannot form a Dirac-Kerr-Newman ring either.
In GR, we do not have (\ref{Pap22}) and (\ref{Pap29}), so a Dirac field can form singular structures lacking spatial extension in more than 1 dimension, such as a Dirac-Kerr-Newman ring as in Burinskii's model \cite{Bur}.
It is possible because these structures do not contradict the invariance of the matter Lagrangian density under tetrad rotations and thus they do not contradict the gravitational field equations.

Since torsion in the ECKS theory prevents Dirac fields from forming point or string configurations, it also determines the minimal spatial extension $d$ of a spinor particle represented by such a field.
The size of this extension is given by the condition at which torsion introduces significant corrections to the energy-momentum tensor, i.e. when the two terms on the right-hand side of (\ref{EC}) are on the same order.
Equivalently, this size is determined by the condition at which the repulsive four-fermion self-interaction Lagrangian term (\ref{HI}) \cite{Ker} balances the gravitationally attractive mass term in (\ref{sECg2}).
The metric energy-momentum tensor for the Lagrangian density (\ref{sECg2}) is on the order of $mc^2|\psi|^2$ (in the rest frame of the particle), the spin density (\ref{sECg4}) is on the order of $\hbar c|\psi|^2$, and the wave function $\psi\sim d^{-3/2}$.
Thus this size is on the order of the Cartan radius $r_C$ \cite{Tr}:
\begin{equation}
\frac{m}{r_C^3}\sim\frac{G}{c^4}\biggl(\frac{\hbar}{r_C^3}\biggr)^2.
\label{order}
\end{equation}
For an electron, $r_{Ce}\sim10^{-27}\,$m, which is 5 orders of magnitude smaller than the currently strongest experimental upper limit on its radius, observed in a Penning trap, $r_e<r_0\sim10^{-22}\,$m \cite{rad}.\footnote{
The experimental upper limit on the size of an electron is much bigger than the radius of the corresponding Kerr-Newman ring.
However, a Kerr-Newman ring describes a free electron.
An electron confined in a Penning trap is subject to strong external fields which can change its Kerr-Newman-like structure into a structure with a spatial extension below $r_0$.
}
It is also much smaller than its Compton wavelength $\frac{h}{mc}\sim10^{-12}\,$m.
For all known fermions, $r_C$ is between $10^{-29}$ and $10^{-25}\,$m.
A discovery of some structure of fermions in this region would be an indirect indication that torsion is different from zero.
We also expect that the quantum-field-theory concept of a fermion (point particle with spin $\frac{1}{2}\hbar$) must be modified if the ECKS theory is correct, introducing an effective ultraviolet cutoff in quantum field theory at distances on the order of $r_{Ce}$.
If GR is correct, an ultraviolet cutoff in quantum field theory would occur at the much smaller Planck scale.

A Dirac field in the ECKS gravity cannot form a singular Dirac-Kerr-Newman ring as in Burinskii's model in GR \cite{Bur} because such a ring lacks a spatial extension along the $r$ and $z$ coordinates.
Since this extension must be on the order of corresponding Cartan radius, our results suggest that the Dirac wave function of an electron acquires a {\em nonsingular} spacetime structure of a {\em toroid} with the outer radius of the electron Compton size and the inner radius of its Cartan size.\footnote{
A toroid seems to be a natural choice because its annular shape resembles that of a ring.
Dirac fermions could possibly form shells of matter, as in \cite{Lo}, although such configurations do not have the same natural physical interpretation as a Kerr-Newman ring does.
}
Note that this suggestion is valid for both charged and uncharged leptons (since $q$ does not affect the derivation of (\ref{der}) and the radii of the torus), and that the weak interaction does not introduce any significant changes.
The toroid description should also work for quarks which are asymptotically free (with respect to the strong interaction) at distances on the order of their Cartan radii.
To verify if a toroidal structure describes a fermion, and also if spacetime surrounding the toroid admits unphysical closed timelike curves as in the case for a Kerr-Newman naked singularity, the full Einstein-Maxwell-Yang-Mills-Dirac-Heisenberg-Ivanenko field equations corresponding to the ECKS Lagrangian density (\ref{ecks}) with $\mathcal{L}_m$ given by (\ref{sECg2}) must be solved.

\section{Discussion}

We showed that free fermions in the ECKS theory of gravity must be extended in at least 2 spatial dimensions and at least on the order of their Cartan radii.
They cannot form point or string distributions because of the conservation law for the spin density.
Such distributions of matter are already problematic in GR (no satisfactory mathematical framework) because it is difficult to find, due to a nonlinear character of the Einstein equations, a class of metrics whose curvature tensors are well-defined as distributions on such submanifolds \cite{Bru,GT}.
Since the ECKS theory may be regarded as GR in which the energy-momentum tensor acquires a correction (\ref{corr}) from the spin of matter, we would expect this theory to have the same problem with distributional curvature for points or strings of matter.
This issue needs further investigation.

We expect that the Cartan density for an electron, $\rho_{Ce}\sim m_e/r_{Ce}^3\sim10^{51}\,\mbox{kg}\,\mbox{m}^{-3}$, approximately gives the order of the maximum density of ordinary matter composed of quarks and leptons.
This limit appears because a system of elementary Dirac particles cannot be compressed to densities higher than the densities of its components (which are on the order of the their Cartan densities), otherwise the particles themselves would be compressed more than it is allowed by (\ref{order}).
Also, the spin density is an additive quantity, so for a system of Dirac spinors it is totally antisymmetric, as for one spinor, leading to (\ref{Pap29}).
Thus we conjecture that singularities are avoided in the ECKS theory for fermions in self-gravitating systems as they are avoided for test Dirac particles.
Gravitational collapse of any configuration of such matter cannot create a singularity, even if an event horizon forms.
Since supernova remnants have mass densities much smaller than the Cartan density for an electron (and for the other fermions), deviations of the ECKS theory from GR are negligible in the evolution of stars and torsion does not prevent the formation of black holes.

Our results generalize previous findings, that spin and torsion can avert cosmological singularities for certain spin configurations \cite{Kop}, to all configurations of matter with spin.
It also agrees with \cite{Hehl2}, which showed that those models violate an energy condition of a singularity theorem.
The mass density of a black hole also cannot exceed $\rho_{Ce}$, from which its minimum mass in the ECKS theory is $\sim10^{16}\,\mbox{kg}$, corresponding to energy $\sim10^{43}\,\mbox{GeV}$.
Therefore the Large Hadron Collider (LHC), which can operate at energies up to $\sim10^{4}\,\mbox{GeV}$, cannot produce micro black holes \cite{Ward} if the four-dimensional ECKS theory is a correct theory of gravity.
In GR, where torsion is absent, a theoretical minimum mass of a black hole is much smaller, near the Planck mass $\sim10^{-9}\,\mbox{kg}$ \cite{Har}.

In deriving (\ref{Pap29}) and showing that it prevents a Dirac field in the ECKS theory from collapsing to a point or system of points we did not use that (\ref{ecks}) is proportional to the Ricci scalar.
We only used that the spin density of such a field is totally antisymmetric.
Thus for other theories of gravity with torsion, (\ref{Pap29}) requires that Dirac particles be extended objects for the same reason as in the ECKS theory as long as the totally antisymmetric component of the torsion tensor couples to spinor fields and the other components do not.
Those theories will likely give different predictions for the size of fermions because of a different condition replacing (\ref{order}).



\begin{thebibliography}{}
\bibitem{LL} L. D. Landau and E. M. Lifshitz, {\em The Classical Theory of Fields} (Pergamon, 1975).
\bibitem{MTW} C. W. Misner, K. S. Thorne, and J. A. Wheeler, {\em Gravitation} (Freeman, 1973); C. M. Will, {\em Theory and Experiment in Gravitational Physics} (Cambridge Univ. Press, 1992).
\bibitem{KN} E. T. Newman, E. Couch, K. Chinnapared, A. Exton, A. Prakash, and R. Torrence, J. Math. Phys. {\bf 6}, 918 (1965).
\bibitem{Car} B. Carter, Phys. Rev. {\bf 174}, 1559 (1968).
\bibitem{KS} G. C. Debney, R. P. Kerr, and A. Schild, J. Math. Phys. {\bf 10}, 1842 (1969).
\bibitem{Isr} W. Israel, Phys. Rev. D {\bf 2}, 641 (1970); A. Y. Burinskii, Russ. Phys. J. {\bf 17}, 1068 (1974); D. Ivanenko and A. Y. Burinskii, Russ. Phys. J. {\bf 18}, 721 (1975); D. Ivanenko and A. Y. Burinskii, Russ. Phys. J. {\bf 21}, 932 (1978); C. L. Pekeris and K. Frankowski, Phys. Rev. A {\bf 39}, 518 (1989); A. Y. Burinskii, Phys. Lett. A {\bf 185}, 441 (1994); A. Burinskii, Phys. Rev. D {\bf 68}, 105004 (2003).
\bibitem{Bur} H. I. Arcos and J. G. Pereira, Gen. Rel. Grav. {\bf 36}, 2441 (2004); A. Burinskii, Phys. Rev. D {\bf 70}, 086006 (2004); A. Burinskii, Grav. Cosmol. {\bf 14}, 109 (2008); A. Burinskii, J. Phys. A {\bf 41}, 164069 (2008).
\bibitem{Lord} E. A. Lord, {\em Tensors, Relativity and Cosmology} (McGraw-Hill, 1976); N. J. Pop{\l}awski, arXiv:0911.0334.
\bibitem{Hehl1} F. W. Hehl, P. von der Heyde, G. D. Kerlick, and J. M. Nester, Rev. Mod. Phys. {\bf 48}, 393 (1976); F. W. Hehl, J. D. McCrea, E. W. Mielke, and Y. Ne'eman, Phys. Rep. {\bf 258}, 1 (1995).
\bibitem{SH} I. L. Shapiro, Phys. Rep. {\bf 357}, 113 (2002); R. T. Hammond, Rep. Prog. Phys. {\bf 65}, 599 (2002).
\bibitem{Tr} A. Trautman, in: J.-P. Francoise, G. L. Naber, and S. T. Tsou (Eds.), {\em Encyclopedia of Mathematical Physics}, vol. 2, p. 189 (Elsevier, 2006).
\bibitem{dS} V. de Sabbata and M. Gasperini, {\em Introduction to Gravitation} (World Scientific, 1986); V. de Sabbata and C. Sivaram, {\em Spin and Torsion in Gravitation} (World Scientific, 1994).
\bibitem{Hehl2} F. W. Hehl, P. von der Heyde, and G. D. Kerlick, Phys. Rev. D {\bf 10}, 1066 (1974).
\bibitem{CF} S. M. Carroll and G. B. Field, Phys. Rev. D {\bf 50}, 3867 (1994).
\bibitem{test} A. S. Belyaev, I. L. Shapiro, and M. A. B. do Vale, Phys. Rev. D {\bf 75}, 034014 (2007).
\bibitem{Pap} M. Mathisson, Acta Phys. Polon. {\bf 6}, 167 (1937); A. Papapetrou, Proc. Roy. Soc. London A {\bf 209}, 248 (1951).
\bibitem{NSH} K. Nomura, T. Shirafuji, and K. Hayashi, Prog. Theor. Phys. {\bf 86}, 1239 (1991).
\bibitem{also} D. Puetzfeld and Y. N. Obukhov, Phys. Rev. D {\bf 76}, 084025 (2007); D. Puetzfeld and Y. N. Obukhov, Phys. Rev. D {\bf 79}, 069902 (2009) (Erratum); D. Puetzfeld and Y. N. Obukhov, Phys. Lett. A {\bf 372}, 6711 (2008); D. Puetzfeld and Y. N. Obukhov, Phys. Lett. A {\bf 373}, 1600 (2009) (Erratum).
\bibitem{IZ} C. Itzykson and J.-B. Zuber, {\em Quantum Field Theory} (Dover, 2005).
\bibitem{antis} O. Costa de Beauregard, Phys. Rev. {\bf 129}, 466 (1963).
\bibitem{HP} H. H\"{o}nl and A. Papapetrou, Z. Phys. {\bf 112}, 512 (1939); A. Papapetrou and H. H\"{o}nl, Z. Phys. {\bf 114}, 478 (1939); H. H\"{o}nl and A. Papapetrou, Z. Phys. {\bf 116}, 153 (1940).
\bibitem{Cor} H. J. Bhabha and H. C. Corben, Proc. Roy. Soc. London A {\bf 178}, 273 (1941); H. C. Corben, Phys. Rev. D {\bf 121}, 1833 (1961); H. C. Corben, {\em Classical and Quantum Theories of Spinning Particles} (Holden-Day, 1968).
\bibitem{BL} R. H. Boyer and R. W. Lindquist, J. Math. Phys. {\bf 8}, 265 (1967).
\bibitem{Hen} R. C. Henry, Astrophys. J. {\bf 535}, 350 (2000).
\bibitem{Mol} C. M{\o}ller, {\em The Theory of Relativity} (Oxford Univ. Press, 1972).
\bibitem{Ker} G. D. Kerlick, Phys. Rev. D {\bf 12}, 3004 (1975).
\bibitem{rad} H. Dehmelt, Phys. Scr. {\bf T22}, 102 (1988).
\bibitem{Lo} C. A. L\'{o}pez, Phys. Rev. D {\bf 30}, 313 (1984).
\bibitem{Bru} A. H. Taub, Illinois J. Math. {\bf 1}, 370 (1957); Y. Four\`{e}s-Bruhat, Compt. Rend. {\bf 248}, 1782 (1959); Y. Bruhat, in: L. Witten (Ed.), {\em Gravitation: An Introduction to Current Research}, p. 130, (Wiley, 1962).
\bibitem{GT} R. Geroch and J. Traschen, Phys. Rev. D {\bf 36}, 1017 (1987); R. M. Wald, arXiv:0907.0412.
\bibitem{Kop} W. Kopczy\'{n}ski, Phys. Lett. A {\bf 39}, 219 (1972); W. Kopczy\'{n}ski, Phys. Lett. A {\bf 43}, 63 (1973); A. Trautman, Nature (Phys. Sci.) {\bf 242}, 7 (1973); J. Stewart and P. H\'{a}j\'{i}\v{c}ek, Nature (Phys. Sci.) {\bf 244}, 96 (1973).
\bibitem{Ward} B. F. L. Ward, J. Cosm. Astropart. Phys. {\bf 02}, 011 (2004).
\bibitem{Har} T. Harada, Phys. Rev. D {\bf 74}, 084004 (2006).
\end{thebibliography}
\end{document}